\documentclass[12pt,preprint]{aastex}

\shorttitle{The braking index of PSR~J1734$-$3333}
\shortauthors{Espinoza et al.}

\begin{document}

\title{The braking index of PSR~J1734$-$3333 and the magnetar population}

\author{C. M.~Espinoza}
\affil{Jodrell Bank Centre for Astrophysics, School of Physics and Astronomy, 
University of Manchester, Manchester, M13~9PL, UK}
\email{cme@jb.man.ac.uk}

\author{A. G.~Lyne}
\affil{Jodrell Bank Centre for Astrophysics, School of Physics and Astronomy, 
University of Manchester, Manchester, M13~9PL, UK}

\author{M. Kramer}
\affil{MPI f\"ur Radioastronomie, Auf dem H\"ugel 69, 53121 Bonn, Germany}

\author{R. N. Manchester}
\affil{Australia Telescope National Facility, CSIRO Astronomy and Space Science,
PO~Box~76, Epping NSW~1710, Australia}

\and

\author{V. M. Kaspi}
\affil{Department of Physics, McGill University, 3600 University Street, 
Montreal, QC H3A 2T8, Canada}



\begin{abstract}
PSR~J1734$-$3333 is a radio pulsar rotating with a period $P=1.17$ s
and slowing down with a period derivative $\dot{P}=2.28\times10^{-12}$, 
the third largest among rotation-powered pulsars.
These properties are midway between those of normal rotation-powered 
pulsars and magnetars, two populations of neutron stars that are 
notably different in their emission properties.
Here we report on the measurement of the second period derivative
of the rotation of PSR~J1734$-$3333 and calculate a braking index 
$n=0.9\pm0.2$.
This value is well below 3, the value expected for an electromagnetic
braking due to a constant magnetic dipole,
and indicates that this pulsar may soon have the rotational properties 
of a magnetar. 
While there are several mechanisms which could lead to such a low braking 
index, we discuss this observation, together with the 
properties exhibited by some other high-$\dot{P}$ rotation-powered pulsars, 
and interpret it as evidence of a possible evolutionary route for magnetars 
through a radio-pulsar phase, supporting a unified description of the two 
classes of object.  
\end{abstract}


\section{Introduction}
Radio pulsars and magnetars are believed to be neutron stars
which have been formed in the collapse of the cores of massive stars 
in supernova explosions  \citep{pac67,dt92a}.
Despite this common origin, they have very different rotational and 
radiative properties 
 \citep{wt06,mer08}.
While radio pulsars generally rotate with periods ${P}$ of $0.01$--$1$ 
second and exhibit no remarkable X-ray emission, magnetars have periods 
ranging from $2$--$12$ seconds and are generally intense sources of 
X-rays, sporadically undergoing outburst episodes.

The original theory of magnetars \citep{dt92a} suggested that some neutron 
stars evolve under different circumstances to normal, or rotation-powered,
radio pulsars.  Convection and fast rotation in their early stages
would quickly generate strong magnetic fields, ultimately producing
Soft Gamma Repeaters (SGRs) and Anomalous X-ray pulsars (AXPs), the two
manifestations of magnetars \citep{wt06}.  Decay of their extremely-high
magnetic field is believed to be the key to explaining outburst episodes 
and the excess of X-ray luminosity over their rate of loss of rotational 
energy  \citep{td95,wt06,mer08}.  

However, it is not clear whether they are indeed born with different 
properties or whether there is an evolutionary relationship between 
the two species \citep{lyn04,km05,ggg+08,hgc+09}.
The discovery of radio pulsars with rotational properties
approaching those of magnetars and the existence of objects of both 
types having some similar emission properties
is now well established  \citep{ksj+07,ggg+08,crj+08,lbb+10}.
We have carried out timing observations of one of these intermediate 
objects, PSR~J1734$-$3333, in order to study its rotational evolution.
In this letter we present the measurement of the braking index of this
pulsar, a most unusual value which defies the standard models for 
pulsar spin-down.

\subsection{Braking index and movement on the $P- \dot{P}$ diagram}
\label{intro2}
The rotational relationship between radio pulsars and magnetars is best
demonstrated in the period -- period-time-derivative ($P- \dot{P}$) 
diagram (Fig.~\ref{fg:ppdot}). 
Radio pulsars are believed to be formed in the upper left region of the
diagram, with short period and large slow-down rate, and move generally
to the right and downwards.  Magnetars have longer periods and larger
slow-down rates than young radio pulsars and are found in the upper 
right region of the diagram.
The chief rotational slow-down mechanism is often assumed to be loss of 
energy through electromagnetic radiation produced by a dipolar rotating 
neutron-star magnetic field. 
Thus, the position in the diagram is conventionally interpreted in 
terms of the constant surface dipolar magnetic field at the magnetic 
equator required to slow down a standard neutron star 
($B_s=3.2\times10^{19}\sqrt{P\dot{P}}$~G), as well as the characteristic 
age $\tau_c=-\nu/2\dot{\nu} = P/2\dot{P}$ of the neutron star (with
$\nu=1/P$)\citep[e.g.][]{ls06}. 
According to this, magnetars have surface magnetic fields of
$10^{14-15}$\,G, about two orders of magnitude higher than those of 
most normal radio pulsars, and are as young as young radio pulsars 
(1--100 kyr).

Motion in the $P-\dot{P}$ diagram can be described by the braking
index, $n$, defined by 
\begin{equation}
\label{powerLaw}
\dot\nu=-k\nu^n \, \quad \mbox{or} \quad \dot{P}=kP^{2-n} \quad ,
\end{equation}
where $k$ is a positive constant.  
With this definition, the slope of the evolutionary path in the diagram 
is equal to $2-n$.  
For pure magnetic braking with a constant dipolar magnetic field 
we expect $n=3$.
As a result, neutron stars are expected to follow lines of constant 
dipolar magnetic field (Fig.~\ref{fg:ppdot}), with a slope of $-1$.
However, systematic variations of the moment of inertia or magnetic 
field, non-dipolar braking resulting from higher-order 
magnetic multipoles or stellar winds could effectively change the 
slope of movement in the diagram  \citep{mdn85,br88}.

Observationally, the braking index can be determined from measurement
of the first two derivatives of $\nu$, since differentiation of the
above power-law (Eq.~\ref{powerLaw}) gives $n={\nu
\ddot{\nu}}/{\dot{\nu}^2}$.  Unfortunately, few values of $n$ have
been determined because young pulsars are often not stable rotators and
short-term timing irregularities in the form of glitches and timing
noise often mask any systematic motion in the $P-\dot{P}$ diagram.
However, with long observational time-baselines it is often possible to
establish the underlying movement across the diagram.  So far, seven
pulsars have steady enough rotation that stable values of braking
index have been determined \citep{lps93,lpgc96,mmw+06,lkg+07,wje11}
(see Table \ref{ta:brakings}).
These values are all less than three and hence the slope of
their movement in the $P$-$\dot{P}$ diagram is somewhat
more positive than $-1$.  
The simple dipolar spin-down model is clearly imperfect.

\section{The braking index of PSR J1734$-$3333}
\label{sct:calculation}
PSR J1734$-$3333 lies in the region between young pulsars and magnetars
on the $P-\dot{P}$ diagram and is a young radio pulsar ($\tau_c=8$ kyr), 
possibly associated with the supernova remnant (SNR) G354.8$-$0.8 
\citep{mbc+02}.  
It has a high inferred dipolar magnetic field ($5 \times 10^{13}$ G), 
one of the highest known values amongst rotation-powered pulsars, 
and similar to the lowest values amongst magnetars (even without
considering the SGR with a very low $\dot{P}$ reported by \citet{ret+10}; 
see Fig.~\ref{fg:ppdot}).  
It was discovered during the Parkes Multibeam Pulsar Survey \citep{mhl+02}, 
and it has been observed regularly since August 1997 by the 64-m 
telescope at Parkes and the 76-m Lovell telescope at Jodrell Bank.
Like other high-$\dot{P}$ radio pulsars, PSR~J1734$-$3333 presents modest
X-ray luminosity, not surpassing its rotational spin-down energy
loss rate  \citep{oklk10}, although with somewhat higher temperature 
than is observed for pulsars with lower effective dipolar magnetic fields
but similar characteristic age  \citep{zkm+10}.
It exhibits timing noise typical of a pulsar of its age 
and has not glitched in 13.5 years, making possible a measurement 
of the $\ddot{\nu}$ due to secular slow-down.  

\begin{figure}
\epsscale{1.0}
\plotone{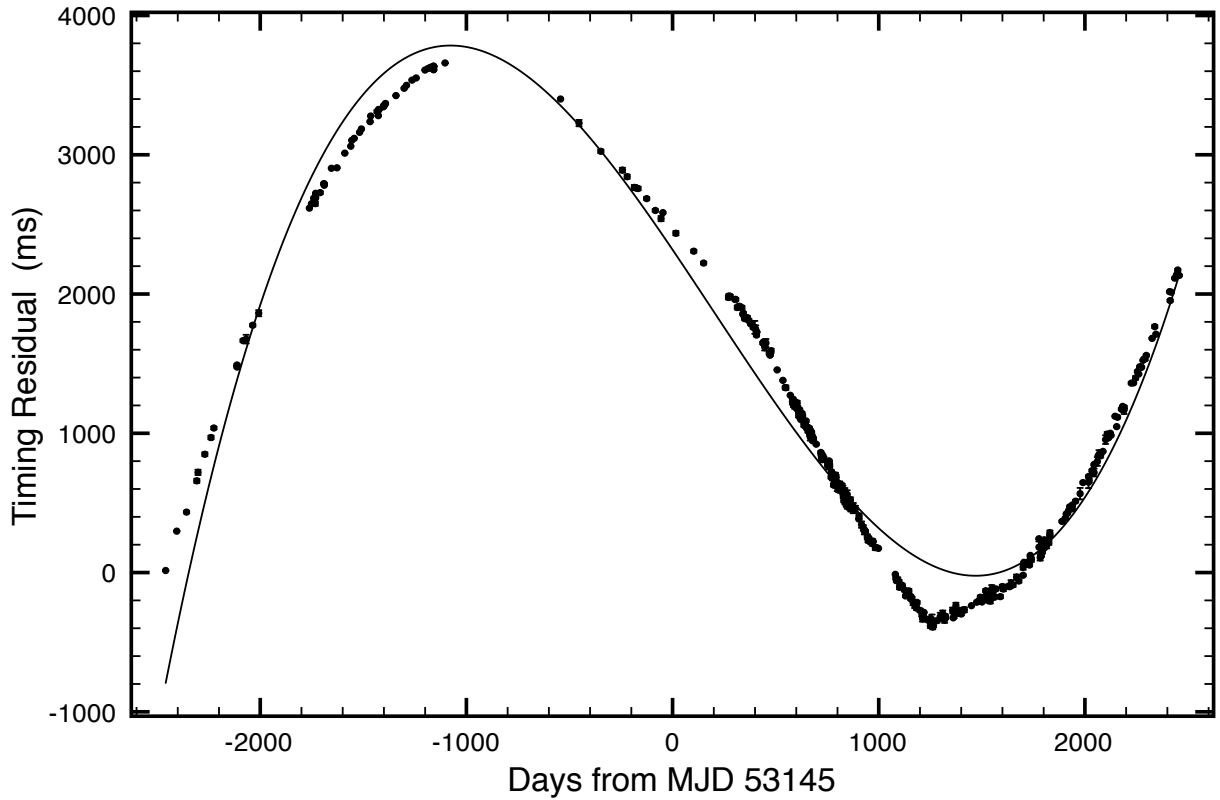}
\caption[]{Timing residuals of PSR~J1734$-$3333 over a 13.5-year
interval. The timing residuals are the difference between the observed
times of arrival of pulses and those predicted by a simple
slow-down model of frequency and first derivative $\dot{\nu}$. 
The curve shows the slow-down model after including the fitted value of frequency
second derivative of $2.8\pm0.6\times10^{-24}$ Hz s$^{-2}$
(Table~\ref{ta:1734pars}), corresponding to a braking index of
$n=0.9\pm0.2$.}
\label{fg:residuals}
\end{figure}

The evolution of the timing residuals of PSR~J1734$-$3333 
relative to a simple slow-down model of the rotational frequency and 
first derivative is shown in Fig.~\ref{fg:residuals}.  
The curve is a fit for a second derivative as well, and these values 
are given in Table~\ref{ta:1734pars}.
In order to assess the errors in the parameters we have measured 
the spectrum of the residuals from the second-derivative fit and 
conducted Monte Carlo 
simulations by randomly varying the phases of the spectral components.
Quoted uncertainties correspond to the standard deviations around the 
mean values.
We obtain $\ddot{\nu}=2.8\pm0.6 \times 10^{-24}$
Hz s$^{-2}$, which implies that the present braking index of this
pulsar is very low, being $n=0.9\pm0.2$.

Many young pulsars such as the Vela pulsar (PSR~B0833$-$45),
PSR~B1800$-$21 and PSR B1823$-$13 have large positive values
$\ddot{\nu}$ associated with recovery from glitches \citep{ls06}.  If
there has been any recent glitch activity in PSR~J1734$-$3333, the
present value of $\ddot{\nu}$ would very likely be contaminated by any such
relaxation and the long-term value would be even smaller \citep{hlk10},
indicating that the actual long-term braking index of PSR~J1734$-$3333
must be less than or equal to $0.9\pm0.2$.

\section{Discussion}
This braking index measurement indicates a movement in the $P$--$\dot{P}$ 
diagram with a slope $2-n$ of at least $0.9$.
The inset diagram in Fig.~\ref{fg:ppdot} shows how, superposed on timing 
noise, there is a clear and systematic motion over nearly 14 years 
towards the top-right region 
of the diagram, where magnetars are found.
If this behavior is sustained, we can estimate the time it will take for 
the pulsar to move from its present position in the diagram to the magnetar 
region.
Using the power law that defines the braking index (Eq. \ref{powerLaw}),
an object relocates in Fig.~\ref{fg:ppdot} from
($P_1,\dot{P}_1$) to ($P_2,\dot{P}_2$), where
$\dot{P_2}=\dot{P_1}(P_2/P_1)^{2-n}$, in a time given by
\begin{eqnarray}
\label{dt}
\Delta T = &
 \frac{2\tau_c}{n-1}\left[\left(\frac{P_2}{P_1}\right)^{n-1}-1\right] &
 ( n\neq 1) \\
\Delta T = &
 2\tau_c {\rm ln}\left(\frac{P_2}{P_1}\right) &
 ( n=1), \nonumber
\end{eqnarray}
where $\tau_c=P_1/2\dot{P}_1$ is the present characteristic age.  
The braking index of a neutron star is determined by the mechanism 
which provides the slow-down torque, due either to electromagnetic radiation 
or particle flow, and any variation in the moment of inertia or magnetic 
field \citep{mdn85,br88,ls06}.
If these underlying physical processes remain unchanged and
PSR~J1734$-$3333 continues to evolve with a constant braking index of
$n=0.9$, Eq.~\ref{dt} implies that it would reach a period of 8~s in only
29~kyr.  It would then be situated in the middle of the magnetars.
The characteristic age would be 6.7~kyr, less than its present 
value of 8.1~kyr and much smaller than the combined durations of its 
previous life as a radio pulsar and its subsequent path amongst magnetars.

Assuming the relation between the neutron star magnetic field and slow-down 
rate given in section \ref{intro2}, the position in the $P$--$\dot{P}$ 
diagram indicates the effective dipole magnetic field strength. 
Most quoted magnetic field strengths, including those of magnetars, are 
based upon this calculation.
The upward movement on the $P$--$\dot{P}$ diagram of PSR~J1734$-$3333 then 
corresponds to an increase of the effective surface neutron-star dipole 
magnetic field.
In this basic model (magnetic-dipole radiation of an orthogonal
rotator in vacuum), $n<3$ can be obtained by variations of either the 
moment of inertia of the star or the magnetic moment \citep{br88}.
While the moment of inertia is indeed expected to decrease in young 
pulsars due to re-shaping caused by the secular spin-down, it is unlikely
that this process can be sustained over long times. 
In contrast, formation of cracks in the crust are expected, that may 
cause glitches or abrupt spin irregularities which would be visible in 
the timing residuals \citep{bppr69,accp96}.
Nothing like this is visible in the data for PSR~J1734$-$3333 during
the 13.5 years of observations.
Variations of the magnetic moment could be caused by alignment or 
misalignment of the magnetic dipole with respect to the spin axis or by 
increase or decrease of the magnetic field strength.
To obtain $n\sim0.9$ the magnetic axis would have to migrate away from 
the rotation axis with a time-scale of $10^4$\,yr.
No such changes have ever been measured and the available evidence 
for secular magnetic moment migration in radio pulsars points to 
alignment (rather than misalignment) on time-scales of $10^{6-8}$\,yr 
\citep{tm98,wj08,ycbb10}.
Hence, the only option in the simple magnetic-dipole radiation scenario
to produce $n<3$ is surface magnetic field growth.

The above interpretation almost certainly involves significant 
simplification. 
Observations have shown that magnetospheric processes could 
exert an extra torque on the crust, perturbing the effects of pure 
electromagnetic braking \citep{klo+06,lhk+10}. 
Indeed, out-flowing plasmas could remove enough angular momentum 
from the star to become the dominant spin-down mechanism, thereby 
giving $n\sim 1$.
In this scenario, it has been predicted that strong relativistic winds 
could produce $n=1$ \citep{mic69,mdn85,ac04b,bta+06}.
We note, however, that \citet{oklk10} found no evidence for extended 
emission around PSR J1734$-$3333 in X-ray observations, suggesting the 
absence of a pulsar wind nebula (PWN).

The ideal magnetic-dipole radiation model has been modified in several 
occasions to produce more realistic models, commonly consisting of a
corotating closed-field-line region surrounded by an open-field-line
region.
These models consider different dipole inclination angles and 
incorporate the effects of a plasma filled magnetosphere 
\citep[e.g.][]{mel97,cs06b,lst11}.
In these scenarios the spin-down can not only be regulated by the 
strength of the dipolar magnetic field, but importantly also by the 
dipole inclination angle, the spin rate of the open field lines 
\citep{cs06b}, and the conductivity of the magnetosphere \citep{lst11}.
In these models, a braking index of three is still expected to be the norm, 
while for a braking index $n<3$ it has to be assumed that either 
the co-rotating magnetosphere ends well inside the light cylinder 
\citep{cs06b} or the conductivity is increasing with time \citep{lst11}. 
Both may explain the case of PSR J1734$-$3333, but we remark that the 
large majority of measured braking indices are not three as expected from 
these models but are smaller than three.
In the case studied here, the braking index is much smaller than three, 
so that we also consider the alternative explanation that the dipole 
surface  magnetic field may increase with time. 
It is possible, that a combination of effects contributes to the observed 
spin-down.

Our ignorance of the actual spin-down mechanism for 
PSR J1734$-$3333 could be partly reduced by considering the properties of 
other high-$\dot{P}$ objects.
The three known radio emitting magnetars exhibit unique radio emission 
properties, not generally observed in normal radio pulsars 
 \citep{ksj+07,crj+08,lbb+10}, which have been attributed to the presence of 
strong magnetic fields.
Another object, PSR~J1846$-$0258, is an X-ray pulsar that in 
2006 showed 
magnetar-like activity after more than 6 years of normal rotation-powered 
pulsar behavior  \citep{ggg+08}.
The effective dipolar magnetic field of this pulsar is amongst the 
highest among normal pulsars and its 
radiative properties during the 2006 outburst, as with the three radio 
magnetars, have been related to the effects of a very high surface magnetic 
field.
By studying an unusual braking index decrease observed after the 2006 event, 
\citet{lnk+11} concluded that the effects of particle wind losses 
(if present) are not significant in the rotation of PSR~J1846$-$0258.
Finally, \citet{hck99} showed (in the case of SGR~1806$-$20) 
that even though its rapid slowdown may be caused by stellar winds, a 
high magnetic field may still be necessary to power the typical magnetar
activity.

Thus, provided that the singular properties of high-$\dot{P}$ neutron 
stars, including magnetars, are effectively driven and determined by 
very strong magnetic fields, the movement of PSR J1734$-$3333 on the 
$P$--$\dot{P}$ diagram could be simply understood as a passage to a 
new magnetar life caused by surface magnetic field growth.
The surface magnetic field of a neutron star could increase due to 
outward diffusion of a stronger internal field  \citep{mp96a,gpz99,ho11}.  
Such a field could have been buried by hypercritical accretion of  
material falling back just after the supernova explosion \citep{che89c}.  
In the case of ohmic diffusion, this rate will depend mainly on the 
depth of the particular submersion, with calculations suggesting that 
the inner field will start emerging with typical timescales between $10^2$ 
to $10^6$~yr  \citep{mp96a,gpz99}. 
This is consistent with the time required for PSR~J1734$-$3333 to 
reach full magnetar spin properties.  
It may be that the only modification required to conventional models 
of magnetar formation is a possible delay of perhaps 10-100~kyr in the 
emergence of the magnetic field.
It is possible that PSR J1734$-$3333 has already undergone some transitory 
magnetar activity and by the time it reaches full magnetar rotational 
properties it may have developed the full radiative properties of a 
magnetar.

\section{Conclusions}
The spin evolution of PSR~J1734$-$3333 during the last 13.5 years is 
consistent with a braking index $n=0.9\pm0.2$, which corresponds to a 
systematic movement in the $P$--$\dot{P}$ diagram with a slope of at 
least $+0.9$, arriving within the region of the magnetars after about 
30~kyr.
We discussed that secular evolution of the magnetic-dipole inclination 
and systematic variations of the moment of inertia are unlikely to 
dominate pulsar spin evolution.
Also, X-ray observations of PSR J1734$-$3333 show no evidence for any 
form of PWN \citep{oklk10} which, if present, would be a signature of 
wind activity around this pulsar that could affect its rotation 
leading to a braking index $n<3$ \citep[e.g.][]{bta+06}.
Most realistic models involving electromagnetic torques predict 
$n\sim3$, although could give $n<3$ if the corotating magnetosphere is 
small \citep{cs06b} or if the magnetosphere's conductivity is increasing
with time \citep{lst11}.

\begin{figure}
\epsscale{0.55}
\plotone{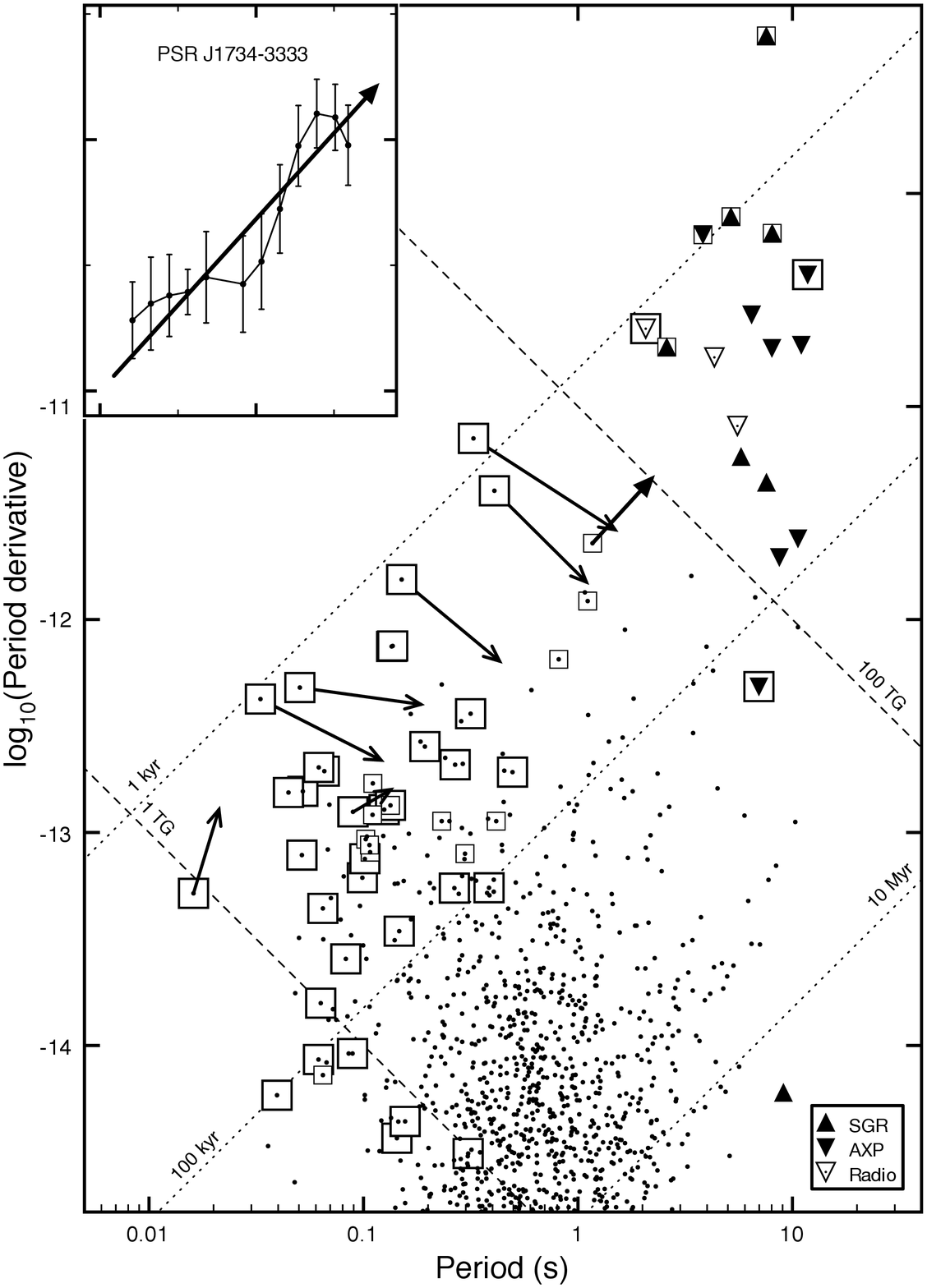}
\caption{$P$--$\dot{P}$ diagram for all known magnetars and
young pulsars having $\dot{P}>1.65\times10^{-15}$. Lines of constant
magnetic field are dashed and lines of constant characteristic age
are dotted, with a slope of 1.  
Open arrows indicate the motion of those pulsars with published
values of braking index (Table~\ref{ta:brakings}).  Each arrow 
represents the projected motion of the pulsar during the next 
$10,000$\,yr (excepting the one for PSR~J0537$-$6910, for which only 
$2,000$\,yr was used), assuming that it evolves with a constant braking 
index $n$.  Equal logarithmic scales are chosen, so that a pulsar will 
move with a slope of 2-$n$.
The motion of PSR~J1734$-$3333 is represented by a closed arrow.
Objects which have robust proposed
associations with supernova remnants or pulsar wind nebulae are identified
with a surrounding large square, with smaller squares for less
convincing associations.
White triangles are used for the radio emitting magnetars. 
The $\dot{P}$ value of the SGR with the lowest $\dot{P}$ corresponds
to an upper limit \citep{ret+10}.
Inset: the pulsar's motion in $P$--$\dot{P}$ space over a 13.5-year
period using equal logarithmic scales, but magnified by a factor of
4,000.
Major ticks are separated by $5\times10^{-4}$~s on the 
horizontal axis, and by $3\times10^{-4}$ on the vertical axis.  
Each data point in the inset is the result of a fit of $P$ and $\dot{P}$
to a 1500-day interval set of times of arrival.
The error bars are the standard deviations (See section \ref{sct:calculation}).
Most of the information was taken from the ATNF pulsar catalogue 
(version 1.39, http://www.atnf.csiro.au/research/pulsar/psrcat/), 
or the McGill SGR/AXP Online Catalog 
as it was in November 2010 
(http://www.physics.mcgill.ca/$\sim$pulsar/magnetar/main.html).
}
\label{fg:ppdot}
\end{figure}

We interpret the movement of PSR J1734$-$3333 in the $P$--$\dot{P}$ 
diagram, together with the properties of other high-$\dot{P}$ neutron 
stars, as evidence for a possible new genesis for at least some magnetars.
Regardless of whether or not this movement corresponds to actual 
magnetic field growth, our interpretation has several
satisfactory implications: 
Firstly, it has been suggested that the apparent excessively high 
birthrate \citep{kk08} of all species of neutron stars over core collapse
supernovae 
events can be explained if there is evolution between the species, as 
we suggest here.  
Secondly, many magnetars will be older than their characteristic 
ages suggest by 10-100~kyr.
Therefore, not all magnetars will be young enough to be still
surrounded by a visible SNR, consistent with the surprisingly small 
number of secure associations between magnetars and SNRs 
(Fig. \ref{fg:ppdot})  \citep{gsgv01,ah04}.
Thirdly, the same physical processes that would be responsible for 
the development of the magnetar properties may be responsible
for the effective dipolar magnetic field growth seen in the
rotation-powered pulsars with measured values of braking index, which 
all have values of $n<3$ (Fig.~\ref{fg:ppdot}, Table~\ref{ta:brakings}).
Finally, recent observations of Central Compact Objects (CCOs) in
SNRs suggest that a significant number of neutron stars are born with 
long periods and low period derivatives  \citep{hg10}.
Hence, young pulsars are not only found at the top-left of 
the diagram but also towards low $\dot{P}$ values, populating 
the left hand side of the main bulk of pulsars and probably being 
progenitors for most normal pulsars.
This may imply that the high-$\dot{P}$ young pulsar population 
are not the progenitors of most pulsars, making feasible their possible 
evolution into magnetars.
In summary, we suggest that, with a range of internal magnetic fields,
submersion depths and initial spin periods, we could explain the 
existence of magnetars, CCOs and the whole radio pulsar population as 
one single family \citep[cf.][]{kas10}.

\acknowledgments
Pulsar research at JBCA is supported by a Rolling Grant from
the UK Science and Technology Facilities Council (STFC).
C.M.E. acknowledges the support received from STFC and CONYCIT through
a PPARC-Gemini fellowship.  V.M.K acknowledges support from NSERC,
FQRNT, CIFAR and from Canada Research and Lorne Trottier Chairs.

\clearpage
\begin{deluxetable}{lcl}
\tablecaption{Published braking indices. 
	\label{ta:brakings}}
\tablehead{
\colhead{Pulsar} & \colhead{$n$} & \colhead{Reference} }
\startdata
B0531$+$21     &  2.51(1)     & \cite{lps93}   \\
J0537$-$6910   &  $-$1.5        & \cite{mmw+06}  \\
B0540$-$69     &  2.140(9)    & \cite{lkg+07}  \\
B0833$-$45     &  1.4(2)      & \cite{lpgc96}  \\
J1119$-$6127   &  2.91(5)     & \cite{wje11}   \\
B1509$-$58     &  2.839(1)    & \cite{lkg+07}  \\
J1846$-$0258   &  2.65(1)     & \cite{lkg+07}  \\
J1734$-$3333   &  0.9(2)      & this work      \\
\enddata
\tablecomments{ The uncertainty in the last shown digit is quoted in 
parenthesis. While all other values come from phase-coherent timing, 
the values for PSR B0833$-$45 and PSR J0537$-$6910 were obtained 
by studying the evolution of the spin-down rate after several glitches. 
The braking index for this last pulsar was calculated from a rough 
estimate of the systematic variation during more than 8 yr of the 
frequency derivative data. We note that the braking index of PSR
1846$-$0258 was found to be $2.2\pm0.1$ when using data following the
2006 event (see section 2) \citep{lnk+11}.}
\end{deluxetable}

\clearpage
\begin{deluxetable}{lc}
\tablecaption{Observed and derived parameters for PSR~J1734$-$3333
obtained from fits to the pulse times-of-arrival.
\label{ta:1734pars} }
\tablehead{
\colhead{Parameter} & \colhead{Value}  }
\startdata
RAJ                        		& 17:34:26.9(2)           \\
DECJ                       		& $-$33:33:20(10)         \\
$\nu$ (Hz)                 	 	& 0.855182765(3)          \\
$\dot{\nu}$ ($10^{-15}$~Hz s$^{-1}$)  	& $-$1667.02(3)           \\
$\ddot{\nu}$ ($10^{-24}~$Hz s$^{-2}$) 	& 2.8(6)      		  \\
$P$ (s)                    		& 1.169340684(4)          \\ 
$\dot{P}$ ($10^{-15})$     		& 2279.41(4)              \\
$\ddot{P}$ ($10^{-24}$~s$^{-1}$)      	& 5.0(8)      		  \\
Timing Epoch (MJD)         		& 53145                   \\
Data span (MJD)            		& 50686-55602             \\
DM (cm$^{-3}$pc)           		& 578(9)                  \\
S$_{1400}$ (mJy)           		& 0.5                     \\
W$_{50}$ (ms)              		& 500                     \\
& \\
Distance (kpc)             		& 7.4                     \\
Characteristic Age (kyr)   		& 8.1                     \\
Surface magnetic field (TG)		& 52                      \\
Braking Index, $n$         		& 0.9(2)                  \\
\enddata
\tablecomments{Standard errors are given in parenthesis in units of 
the last quoted digit. See section \ref{sct:calculation} for more 
details.}
\end{deluxetable}

\end{document}